\documentclass[12pt,a4paper]{article}
\usepackage[latin1]{inputenc}
\usepackage[english]{babel}
\usepackage{amsmath}
\usepackage{amsfonts}
\usepackage{textcomp}
\usepackage{amssymb}
\usepackage{physics}
\usepackage{tensor}
\usepackage{float}
\usepackage{times}
\usepackage{enumerate} 
\usepackage[nottoc,notlot,notlof]{tocbibind}
\usepackage{graphicx}
\usepackage[latin1]{inputenc}
\usepackage[usenames]{color}
\usepackage[left=3.5cm,right=2.5cm,top=2.5cm,bottom=2.5cm]{geometry}
\usepackage{color}
\usepackage{latexsym}
\usepackage{amsmath}
\usepackage{amsfonts}
\usepackage{amssymb}
\usepackage{indentfirst} 
\usepackage{tikz}

\title{Superintegrable dynamics on $H^{2}$ generated by coupling the Morse and Rosen-Morse potentials}
\author{John Acosta\thanks{e-mail: john.acosta@unilodz.eu}, \quad  Cezary Gonera\thanks{e-mail: cgonera@uni.lodz.pl}\\\\
\small Faculty of Physics and Applied Informatics, \\
\small University of \L\'od\'z,\\
\small Pomorska 149/153, 90-236 {\L}\'od\'z, Poland
}
\date{}
\begin{document}
\maketitle 
\begin{abstract}
A Hamiltonian dynamics defined on the two-dimensional hyperbolic plane by coupling the Morse and Rosen-Morse potentials is analyzed. It is demonstrated that orbits of all bounded motions are closed iff the product of the parameter $\tilde{a}$ of the Morse potential and the square root of the absolute value of the curvature is a rational number.
This property of trajectories equivalent to the maximal superintegrability is confirmed by explicit construction of polynomial superconstant of motion.
\end{abstract}

\section{Introduction}
The Morse \cite{Morse} and the Rosen-Morse \cite{Rosen}  potentials were originally introduced to describe the potential energy of diatomic particles.\\
The potentials being analytically solvable have found numerous applications in various branches of physics ranging from cosmology \cite{cosmo} through the supersymmetric quantum mechanics \cite{QM}, condensed matter physics \cite{Cond} including Weyl semimetals \cite{Weyl} representing topological materials.\\
In the present paper a coupling of the Morse potential to the Rosen-Morse one is considered in the context of superintegrability. We introduce a Hamiltonian system with two-dimensional hyperbolic configuration space. In the so-called geodesic parallel coordinates (which go to the standard Cartesian ones of Euclidean plane in the zero curvature limit) the dynamics of the system is separable and effectively described by the Morse and Rosen-Morse potentials. We define conditions on the parameters of these potentials, in particular, on the curvature of the configuration space, which make orbits of all bounded motions closed. It is known that in the case of the system with two degrees of freedom this is equivalent to the existence of three globally defined and functionally independent constants of motion i.e. to a maximal superintegrability of the model (in general, for a system of $n$ degrees of freedom the number of global, functionally independent integrals of motion necessary for maximal superintegrability is $2n-1$, see ref. \cite{Arnold},\cite{Per},\cite{Babelon}, the reviews \cite{Miller}, \cite{Kress} and references therein for appropriate definitions and general settings). Due to the time independence and separability of our model, two global and functionally independent integrals are provided by the Hamiltonian itself and a function of dynamical variables corresponding to the separation constant.\\
Using action-angle variables technique we construct explicitly the third functionally independent and global constant of motion. It appears to be polynomial in momenta what is often referred to as higher-order polynomial superintegrability. This kind of superintegrability, characteristic for anisotropic oscillators with commensurable frequencies both on Eucliden and curved spaces \cite{Ranada},\cite{10},\cite{11},\cite{12},\cite{13},\cite{14},\cite{15},\cite{16},\cite{Negro}, was also observed and attracted some attention in the case of Tremblay-Turbiner-Winternitz (TTW) \cite{TTW},\cite{TTW2}, \cite{TTW3} and Post-Winternitz (PW) \cite{PW}, \cite{PW2} models on Euclidean plane as well as their generalizations to 2D sphere and Hyperbolic plane \cite{Hak},\cite{Biz},\cite{Ran},\cite{Hak2}.\\
Another 2D models allowing for higher-order, not necessarily, polynomial integral have been considered in \cite{pol1},\cite{pol2}\cite{pol3},\cite{pol4},\cite{pol5} and more recently, in a series of articles devoted to the higher-order superintegrrability of systems separating in the polar coordinates on Euclidean plane\cite{l31},\cite{l32},\cite{l33}. In fact, replacing the radial oscillator / Kepler potentials of those Euclidean model by relevant curved radial potentials given in \cite{Gon1}\cite{Gon2} should result in a polynomially superintegrable systems on 2D sphere or hyperbolic plane, respectively. Finally, let us note that our model being polynomially superintegrable, is not, in general, included in the interesting wide class of quadratically superintegrable systems on hyperbolic plane constructed in \cite{Ranada}.
On the other hand, being separable in the parallel geodesic coordinates, it yet does not belong, at least explicitly, to the families of the superintegrable models defined on hyperbolic plane and separable in a polar geodesic coordinates described in \cite{Gon1},\cite{Gon2}.

\section{Superintegrable dynamics on hyperbolic plane generated by a coupling of the Morse and Rosen-Morse potentials}

We consider a dynamical system on two-dimensional hyperbolic coordinate space $H^{2}$. In so-called geodesic parallel coordinates $x,y$ (which go to the usual Cartesian ones on Euclidean plane, in zero curvature limit, see for example\cite{Ranada})it is defined by a Hamiltonian

\begin{equation}
\label{1}
H(x, y, p_{x}, p_{y})= H_{0}(y, p_{x}, p_{y})+V(x,y)
\end{equation}
\\
where a free, kinematic part of $H$ (determined by metric $ds^2=\cosh^2(\sqrt{\abs{k}})dx^2+dy^2$ of $H^{2}$ in the $x,y$ variables) reads,

\begin{equation}
\label{2}
H_{0}(y, p_{x}, p_{y})=\frac{1}{2m}(p_{y}^2+\frac{p_{x}^{2}}{\cosh^2(\sqrt{\abs{k}}y)});
\end{equation}
\\
here, $p_x$ and $p_y$ are canonical momenta conjugated to $x,y$ coordinates, respectively,\\
$m>0$ is a positive number standing for mass and,\\
$k= -\abs{k}$ is the curvature of the hyperbolic plane.\\
A potential $V(x,y)$ governing the dynamics is given by the formula

\begin{equation}
\label{3}
V(x,y)=\frac{U(x)}{\cosh^2(\sqrt{\abs{k}}y)}+\Omega\tanh(\sqrt{\abs{k}}y)
\end{equation}
\\
where $\Omega > 0$ is a positive number, while $U(x)$ is the Morse potential,

\begin{equation}
\label{4}
U(x)=V_0(e^{-x/\tilde{a}}-1)^2-V_0 \qquad \tilde {a},  V_0>0 ;
\end{equation}
\\
$\tilde {a}$ and $V_0$ are the parameters of the potential controlling its width and depth, respectively (see Fig.1)
\begin{figure}[H]
\centering
\includegraphics[scale=0.7]{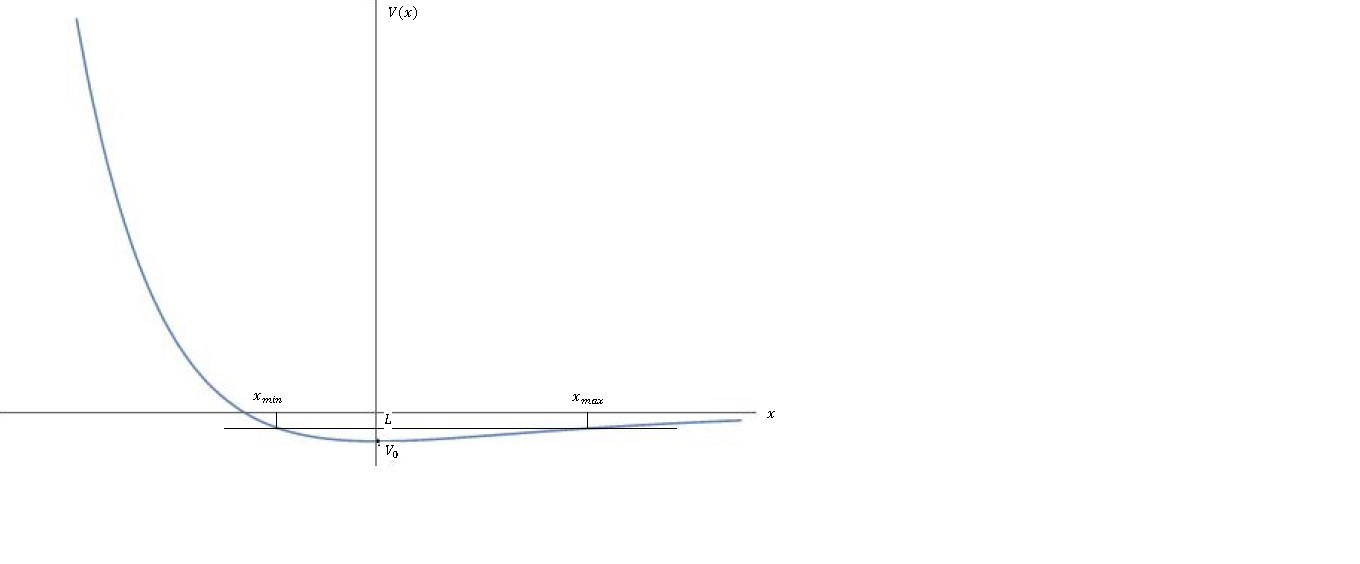}
\caption{The Morse potential}
\end{figure}

Note that if $U(x)$ were a constant; $U(x)=V_{0}$ one would deal with the Rosen-Morse potential (see Fig.2) so the potential $V(x,y)$ can be regarded as coupling the Morse and Rosen-Morse potentials.\\
Due to the form of the $H_{0}$ and $V(x,y)$ the Hamiltonian (1) allows separation of variables.
Indeed, introducing Morse Hamiltonian $H_{M}(x,p_{x})$,

\begin{equation}
\label{5}
H_{M}(x, p_{x})=\frac{p_{x}^2}{2m}+U(x)
\end{equation}
\\
and rewriting the original Hamiltonian (\ref{1}) in the form,
\begin{equation}
\label{6}
H(x, y, p_{x}, p_{y})=\frac{p_{y}^2}{2m}+\frac{H_{M}(x, p_{x})}{\cosh^2(\sqrt{\abs{k}}y)}+\Omega\tanh(\sqrt{\abs{k}}y)
\end{equation}
\\
shows that the Morse Hamiltonian Poisson commuting with H ($\{H,H_{M}\}=0$) provides the global constant of motion $L=H_{M}(x,p_{x})$ separating the dynamics.\\
The time evolution of $x$ coordinate is defined by the Morse Hamiltonian while that of $y$ variable by the original Hamiltonian (6) with $H_{M}(x, p_{x})$ being replaced by L;

\begin{equation}
\label{7}
H(x, y, p_{x}, p_{y})=\frac{p_{y}^2}{2m}+\frac{L}{\cosh^2(\sqrt{\abs{k}}y)}+\Omega\tanh(\sqrt{\abs{k}}y)
\end{equation} 
The potential part of this Hamiltonian is just the Rosen-Morse potential,

\begin{equation}
\label{8}
V_{RM}(y)=\frac{L}{\cosh^2(\sqrt{\abs{k}}y)}+\Omega\tanh(\sqrt{\abs{k}}y)
\end{equation}
\\
Both, the Morse (\ref{3}) and Rosen-Morse (\ref{8}) potentials admit bounded and unbounded motions.
We will be interested in the bounded ones here.\\
For all values of parameters $V_{0}, \tilde{a}>0$ there is a bounded region with a single minimum $U_{min}=-V_{0}$ at $x=0$ in the Morse potential; actually, for negative values of $L=H_{M}(x, p_{x})$ all motions in the $x$ variable are bounded and periodic.\\
On the other hand, an existence of the bounded region with a single minimum $(V_{RM})_{min}=L+\Omega^2/4L$ at $y_{0}$ such that $\tanh(\sqrt{\abs{k}}y_{0})= \Omega/2L$ in the Rosen-Morse potential requires a condition,
 
\begin{equation}
\label{9}
\frac{\Omega}{2}<\abs{L}
\end{equation}
\\
to be met. This, in particular, imposes the relation

\begin{equation}
\label{10}
\Omega<2V_{0}
\end{equation}
\\
for the parameters of the potentials $U(x), V_{RM}(y)$ and together with the behaviour of the Rosen-Morse potential as $y$ tends to $\pm \infty$; $\lim_{y\to \pm \infty}V_{RM}(y)= \pm \Omega$, shows that for the energies from the interval $(L+\Omega ^2/4L, -\Omega)$ the motion in $y$ variable is bounded (see fig.2),

\begin{figure}[H]
\centering
\includegraphics[scale=0.65]{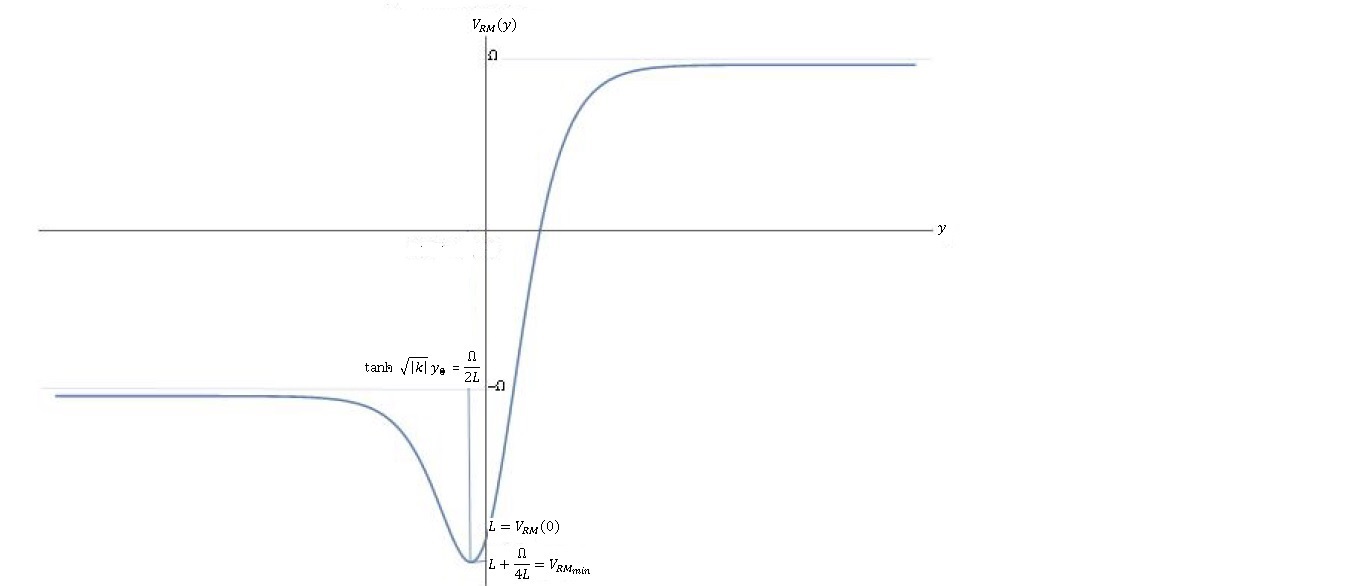} 
\caption{The Rosen-Morse potential}
\end{figure} 

Ultimately, we conclude that the two-dimensional integrable dynamics defined by Hamiltonian (\ref{1}) is bounded if the separation constants $L=H_{M}(x, p_{x}), E=H(x,y,p_{x},p_{y})$ satisfy the following conditions,

\begin{equation*}
L+\frac{\Omega^2}{4L}<E<-\Omega
\end{equation*}
\begin{equation}
\label{11}
-V_{0}<L<0
\end{equation}
\\
Then by Arnold-Liouville theorem equations, 
\begin{equation*}
H(x,y,p_{x},p_{y})=E
\end{equation*}
\begin{equation}
\label{12}
H_{M}(x,p_{x})=L
\end{equation}
\\
define a compact surface in the phase space diffeormorphic to two-dimensional $\textit{Arnold-Liouville}$ tori. This allows to introduce the $\textit{action-angle}$ variables $(I_{x},I_{y},\Psi_{x},\Psi_{y})$ which we will use to derive the condition on the existence of third global and functionally independent on $H$ and $H_M$ constant of motion.\\
In our case, the action variables $I_{x},I_{y}$ are defined as follows,

\begin{equation}
\label{13}
I_{x}=\int_{x_{min}}^{x_{max}}\sqrt{2m(L-U(x))}dx
\end{equation}

\begin{equation}
\label{14}
I_{y}=\int_{y_{min}}^{y_{max}}\sqrt{2m(E-V_{RM}(y))}dy
\end{equation}
\\
where $x_{min/max}, y_{min/max}$ are the roots of the relevant integrands (\ref{13}), (\ref{14}) and $U(x), V_{RM}(y)$ are the Morse and Rosen-Morse potentials given by the equations (\ref{4}) and (\ref{8}) respectively.\\
Performing the integrals yields

\begin{equation}
\label{15}
I_{x}= \tilde{a}\sqrt{2m}(\sqrt{V_0}-\sqrt{-L})
\end{equation}
\\
\begin{equation}
\label{16}
I_{y}= \frac{\sqrt{2m}}{\sqrt{\abs{k}}}(\sqrt{-L}-\frac{1}{2}(\sqrt{-(\Omega+E)}+\sqrt{\Omega-E} ))
\end{equation}
\\
which, in turn, implies

\begin{equation}
\label{17}
I_{y}+\frac{I_{x}}{\tilde{a}\sqrt{\abs{k}}}=  \frac{\sqrt{2m}}{\sqrt{\abs{k}}}(\sqrt{V_{0}}-\frac{1}{2}(\sqrt{-(\Omega+E)}+\sqrt{\Omega-E} )=f(E)   
\end{equation}
\\
Hence, iff the product of the parameter $\tilde{a}$ of the Morse potential and the square root of the absolute value of the curvature is a rational number i.e. $\sqrt{\abs{k}} $  and $1/\tilde{a}$ are commensurable;

\begin{equation}
\label{18}
\tilde{a}\sqrt{\abs{k}} = \frac{r}{s} \qquad  r,s \in \mathbb{Z}
\end{equation}
\\
we obtain

\begin{equation}
\label{19}
rI_{y}+sI_{x}= pf(E)= \tilde{f}(E)
\end{equation}
\\
This crucial relations says that linear combination of the action variables with integer coefficients $r,s$ is the function $\tilde{f}(E)$ of energy only provided $\tilde{a}$ and $k$ satisfy the equation (\ref{18}). Inverting the relation (\ref{19}) one finds

\begin{equation}
\label{20}
H=H(I_{x}, I_{y})= H(sI_{x}+rI_{y})
\end{equation}
\\
This leads to a linear dependence of frequencies $ \frac{\partial{H}}{\partial{I_{y}}}, \frac{\partial{H}}{\partial{I_{x}}}$

\begin{equation}
\label{21}
s\frac{\partial{H}}{\partial{I_{y}}}=r\frac{\partial{H}}{\partial{I_{x}}}
\end{equation}
\\
and, in consequence, to the superintegrability of the system. The additional,functionally independent globally defined constant of motion $C$ can be written as a real or imaginary part of the quantity
 
\begin{equation}
\label{22}
C= F(I_x, I_y)e^{i\phi}
\end{equation} 
\\
where $F$ is any smooth function of actions variables while the non-single valued constant "phase" $\phi$ reads

\begin{equation}
\label{23}
\phi=(r\Psi_{x}-s\Psi_{y}) \qquad r,s \in \mathbb{Z}
\end{equation}
\\
In equation (\ref{23}) $\Psi_{x}, \Psi_{y}$ are the angle variables canonically conjugated to the actions $I_{x}, I_{y}$ respectively.\\
Indeed, $C$ is globally defined and conserved because its time derivative is proportional to the time derivative of phase $\phi$ $(\dot{C}= i\dot{\phi}C)$ which vanishes due to the Hamiltonian equations of motion and the relation (\ref{21}). 

\section{Polynomial superconstant of motion}

It is worth emphasizing here, that an explicit form of the superconstant of motion $C$ is not necessary to prove the superintegrablity of the model. The very existence of $C$ is a consequence of the way the Hamiltonian function depends on the actions.\\
Nevertheless, the explicit form of the superintegral $C$ may provide an useful information concerning certain properties of the system, such as Poisson algebra of integrals of motion or their dependence on the momenta. Therefore, below we present the explicit form of the $C$.\\
The angle variables $\Psi_{x}, \Psi_{y}$ entering the phase $\phi$, given by equation (\ref{23}), are defined as partial derivatives (with respect to the relevant actions) of a function $\tilde{S}(x,y,I_{x},I_{y})$ generating the canonical transformation just to the action-angle variables;

\begin{equation}
\label{24}
\Psi_{x}= \frac{\partial{\tilde{S}}}{\partial{I_{x}}} \qquad \Psi_{y}= \frac{\partial{\tilde{S}}}{\partial{I_{y}}}
\end{equation}
\\
The function $\tilde{S}(x,y,I_{x},I_{y})$ is given by the formula

\begin{equation}
\label{25}
\tilde{S}(x,y,I_{x},I_{y})=S(x,y,E,L)|_{E=E(pI_{y}+qI_{x}), L=L(I_{x})}
\end{equation}
\\
where $S(x,y,E,L)$ reads,

\begin{equation}
\label{26}
S(x,y,E,L)=\sqrt{2m}\int_{y_{min}}^{y}\sqrt{E-V_{RM}(y)}dy+\sqrt{2m}\int_{x_{min}}^{x}\sqrt{L- U(x)}dx
\end{equation}
\\
Hence

\begin{equation}
\label{27}
\phi=r\Psi_{x}-s\Psi_{y}= r\frac{\partial{S}}{\partial{L}}\frac{dL}{dI_{x}}= \frac{2r\sqrt{-L}}{\tilde{a}\sqrt{2m}}\frac{\partial{S}}{\partial{L}}
\end{equation}
\\
In the last step of the above formula, equation (\ref{15}) has been used to compute $dL/dI_{x}$.\\
Performing the relevant integrals gives

\begin{equation}
\label{28}
\phi=s\arcsin (G(y))-r\arcsin (H(x))
\end{equation}
\\
where the functions $G$ and $H$ read, respectively

\begin{equation*}
G(y)=\frac{2L\tanh (\sqrt{\abs{k}}y) -\Omega}{\sqrt{4L^{2}-4LE+\Omega ^{2}}}
\end{equation*}
\begin{equation}
\label{29}
H(x)=\frac{V_{0}e^{-x/\tilde{a}}+L}{e^{-x/\tilde{a}}\sqrt{V_{0}(V_{0}+L)}}
\end{equation}
\\
Inserting the $\phi$ into the equation (\ref{22}) and simplifying the resulting expression by choosing an appropriate function $F(I_{x},I_{y})=F(L,E)$ we find the following explicit form of the additional integral of motion

\begin{equation}
\label{30}
C=e^{\frac{rx}{\tilde{a}}}\left(\sqrt{\frac{-2L}{m}}p_{y}+i(2L\tanh (\sqrt{\abs{k}}y) -\Omega)\right)^{s}\left(\sqrt{\frac{-2L}{m}}p_{x}-i(2V_{0}e^{\frac{-x}{\tilde{a}}}+L)\right)^{r}
\end{equation}
\\
It can be explicitly checked that $C$ is a constant of motion by taking the time derivative of $C$ and using the Hamilton equations of motion as well as the equation (\ref{18}). The real integral of motion is provided by the real or imaginary part of $C$ (obviously, only one of them is functionally independent).\\
Similarly to the TTW model, both real and imaginary part of the $C$ contain either even or odd positive powers of $\sqrt{-L}$, the square root of the separation constant \cite{Gon3}.  Hence multiplying by $\sqrt{-L}$, if necessary, we arrive at a polynomial in the momenta $p_{x}, p_{y}$ constant of motion $C$.\\
As $C(L=0)=(-i)^{r+s}\Omega ^{s} (2V_{0})^{r}$ one can subtract $C(L=0)$ and next divide $C - C(L=0)$ by $L$, obtaining in this way, a polynomial in momenta $p_{x}, p_{y}$ of degree $2(r+s-1)$.

\section{Summary}

We have considered  the Hamiltonian system with two dimensional hyperbolic configuration space. The dynamics of the system separating in the geodesic parallel coordinates is effectively defined by the Morse and the Rosen-Morse potentials.\\
It has been demonstrated that the orbits of all bounded motions are closed iff the parameter $\tilde{a}$ of the Morse potential and the curvature $k$ of the configuration space satisfy the equation (\ref{18}) i.e. when $\sqrt{\abs{k}}$ and $1/\tilde{a}$ are commensurable. It means that,in addition to the Hamiltonian and constant of motion related to the separability, there exists the third global and functionally independent on $H$ and $L$ integral of motion. We have found its explicit form and showed that it is a polynomial in momenta of degree $2(r+s-1)$.\\
A superintegrability of classical system is a fragile property, which in general does not survive a quantization procedure, nevertheless; both the polynomial superintegrability of our classical model and the solvability of its quantized version make us to suppose that the model remain also superintegrable on the quantum level. This will be analyzed elsewhere.\\
Another potentially interesting question is the possible generalization of the model. The answer would consist in specifying all potentials separating in geodesic parallel coordinates and leading to a superintegrable dynamics.

\newpage

\section*{Acknowledgments}

We thank Joanna Gonera and Piotr Kosinski for helpful discussion and comments.\\

J.A. is very grateful to Dr. Katarzyna Stanislawska and the team of nurses for professional and warm medical care extended to him during his stay in the unit of neurosurgery and neurotraumatology of the Heliodor Swiecicki Hospital in Poznan.

\end{document}